\documentclass[10pt,journal]{IEEEtran}
\usepackage{amssymb}
\usepackage{graphicx}
\usepackage{epstopdf}
\usepackage{gensymb}
\usepackage{color}
\usepackage{amsmath}
\usepackage{bm}
\usepackage{etoolbox}
\usepackage{cite}
\usepackage{array}
\usepackage{booktabs}
\usepackage{multirow}
\usepackage{comment}
\usepackage{subcaption}
\usepackage[ruled,vlined]{algorithm2e}
\usepackage{algorithmicx}
\usepackage{mathrsfs}
\usepackage{algpseudocode}
\usepackage{hyperref}

\usepackage{balance}

\usepackage{soul}
\hypersetup{
    colorlinks=true,
    linkcolor=black,
    filecolor=black, 
    citecolor=black,
    urlcolor=cyan,
}

\usepackage[framemethod=TikZ]{mdframed}
 
\alglanguage{pseudocode}

\algnewcommand{\Initialize}[1]{%
  \State \textbf{Initialize:}
  \Statex \hspace*{\algorithmicindent}\parbox[t]{.8\linewidth}{\raggedright #1}
}

\usepackage{threeparttable}
\usepackage{mathtools}
\usepackage{hyperref}
\usepackage{pifont}
\usepackage{listings}

\usepackage{url}
\urldef{\mailsa}\path|{alfred.hofmann, ursula.barth, ingrid.haas, frank.holzwarth,|
	\urldef{\mailsb}\path|anna.kramer, leonie.kunz, christine.reiss, nicole.sator,|
	\urldef{\mailsc}\path|erika.siebert-cole, peter.strasser, lncs}@springer.com|    

\usepackage[utf8]{inputenc}
\usepackage[english]{babel}

\usepackage{amsthm}

\usepackage{soul}
\usepackage{calc}
\newcommand{\textover}[3][l]{%
	\makebox[\widthof{#3}][#1]{#2}%
}

\usepackage{minibox}
\newcounter{observcntr}
\newcommand*{\observ}[1]{%
    \stepcounter{observcntr}%
    \begin{center}
    \vspace{-4pt}
    \minibox[frame, rule=1pt,pad=3pt]{
        \begin{minipage}[t]{0.95\columnwidth}
        \textbf{Observation~\arabic{observcntr}:} \textit{#1}.
        \end{minipage}
    }
    \vspace{-4pt}
    \end{center}
}

\theoremstyle{definition}

\definecolor{R}{RGB}{0,0,150}

\theoremstyle{remark}

\usepackage[framemethod=TikZ]{mdframed}

\ifCLASSINFOpdf

\newcommand{\eat}[1]{}

\begin{document}
\title{Towards A Critical Evaluation of Robustness for Deep Learning Backdoor Countermeasures}

\author{
        Huming Qiu\IEEEauthorrefmark{1},
        Hua Ma\IEEEauthorrefmark{1},
        Zhi Zhang,
        Alsharif Abuadbba,
        Wei Kang,
        Anmin Fu,
        Yansong Gao\IEEEauthorrefmark{2}
	\thanks{H.~Qiu\IEEEauthorrefmark{1} and H.~Ma\IEEEauthorrefmark{1} are joint first authors.}
	\thanks{Y.~Gao\IEEEauthorrefmark{2} is the corresponding author.}
	
	\thanks{H. Qiu, A. Fu. Y.~Gao are with School of Computer Science and Engineering, Nanjing University of Science and Technology, China.
	\{12010622268;fuam;yansong.gao\}@njust.edu.cn}
	\thanks{H.~Ma is with School of Electrical and Electronics Engineering, The University of Adelaide, Australia, and Data61, CSIRO. hua.ma@adelaide.edu.au}
	\thanks{Z.~Zhang, A.~Abuadbba, and W.~Kang are with Data61, CSIRO, Australia. E-mail: \{zhi.zhang;sharif.abuadbba;wei.kang\}@data61.csiro.au}

}

\maketitle

\begin{abstract}
Since Deep Learning (DL) backdoor attacks have been revealed as one of the most insidious adversarial attacks, a number of countermeasures have been developed  with certain assumptions defined in their respective threat models. However, the robustness of these countermeasures is inadvertently ignored, which can introduce severe consequences, e.g., a countermeasure can be misused and result in a false implication of backdoor detection.

For the first time, we critically examine the robustness of existing backdoor countermeasures with an initial focus on three influential model-inspection ones that are Neural Cleanse (S\&P'19), ABS (CCS'19), and MNTD (S\&P'21). Although the three countermeasures claim that they work well under their respective threat models, they have inherent unexplored non-robust cases depending on factors such as given tasks, model architectures, datasets, and defense hyper-parameter, which are \textit{not even rooted from delicate adaptive attacks}. We demonstrate how to trivially bypass them aligned with their respective threat models by simply varying aforementioned factors. Particularly, for each defense, formal proofs or empirical studies are used to reveal its two non-robust cases where it is not as robust as it claims or expects, especially the recent MNTD. This work highlights the necessity of thoroughly evaluating the robustness of backdoor countermeasures to avoid their misleading security implications in unknown non-robust cases.

\end{abstract}

\IEEEpeerreviewmaketitle


\section{Introduction}

Since backdoor attacks first appeared in 2017~\cite{gu2017badnets,chen2017targeted}), they have posed a serious challenge to Deep Learning (DL)-enabled security-sensitive applications such as face recognition~\cite{wenger2021backdoor}, self-driving~\cite{ma2022dangerous}, and medical diagnosis~\cite{nwadike2020explainability}. To mitigate them, significant  efforts have been spent in proposing countermeasures, which can be divided into two categories~\cite{gao2020backdoor} that are data inspection and model inspection, and each inspection can be done offline or online. 
A data-inspection approach audits every input sample either offline or online and decides  whether a given sample is malicious (i.e., containing a trigger) or benign.
A model-inspection method is usually performed  offline, which examines whether a given model is backdoored or not. If yes, a model user either rejects the model before its deployment or fixes it by removing its backdoor effect. Compared to the data inspection, the model inspection often requires DL expertise and/or computing-intensive resources; however,  it can be done only once without further auditing the model or its input samples, making itself preferable.

\subsection{Limitations of State-of-The-Arts} Under predefined threat models, the model inspection can be effective against targeted backdoor types or trigger types (refer to Section~\ref{sec:type}). For example, 
some model-inspection approaches assume that a trigger for a backdoor attack is not large and thus they are effective against common source-agnostic backdoor attacks, where an input sample that is embedded with a small trigger (e.g., a fixed patch) can hijack an infected model.
Yet, adaptive attacks  (e.g., delicately crafted regularization is applied to backdoor a model) can defeat such countermeasures, which are regarded to be challenging to address in the arms-race of DL security~\cite{carlini2019evaluating,carlini2017adversarial,ma2019nic,shokri2020bypassing,gao2021design,tang2021demon,gao2020backdoor}. 

However, we observe that robustness of existing model-inspection approaches has never been critically investigated, e.g., whether the effectiveness of a given defense is dependent on \textit{datasets, tasks, model architectures, model hyperparameters, etc.}. When the defense is used in a non-robust case, it is likely to fail in detecting a backdoor attack even under its predefined threat model (e.g., assuming a source-agnostic backdoor with a small fixed trigger), which \textit{is not even caused by adaptive attacks}. 

To this end, we ask the following questions:



\begin{mdframed}[backgroundcolor=black!10,rightline=false,leftline=false,topline=false,bottomline=false,roundcorner=2mm]
Is a given (influential) backdoor countermeasure (in particular, model inspection) robust enough in different scenarios within its well-predefined threat model? If not, what factors make it non-robust and to which extent render its ineffective in backdoor detection?
\end{mdframed}

The questions above are critical for a user as she can misuse an influential countermeasure without fully understanding its robustness, resulting in a false backdoor detection. That is, even if the countermeasure claims that a model is backdoor-free within its threat model, it does not mean that the model is not backdoored with the defense-targeted trigger or/and backdoor type.

\subsection{Our Work and Contributions} 
This work is, for the first time, towards revealing the non-robust cases of existing backdoor countermeasures, with an initial focus on influential model-inspection solutions (i.e., Neural Cleanse (S\&P'19) \cite{wang2019neural}, ABS (CCS'19)~\cite{liu2019abs}, and MNTD (S\&P'21)~\cite{xu2021detecting}). The main reason why we investigate the model-inspection ones is that they require no access to training data with only one-off offline cost. We throw unknown security caveats of these countermeasures, which indicate that their robustness must be thoroughly examined to avoid any misleading usage. 

Specifically, we reevaluate the robustness of the aforementioned three model-inspection defenses. For each defense, we have revealed two non-robust cases using either formal proofs or empirical case studies and these cases are well aligned with its predefined threat model. Their non-robustness are due to unexplored dependence factors including given tasks, model architectures, datasets, and defense hyper-parameter. We believe that this work can open a new research direction of examining robustness of existing backdoor countermeasures, particularly the influential ones, before they are deployed in real-world. 

The main contributions and results are summarized below---our evaluations are based on reproducing the source code of each original work.

\begin{itemize}
    \item We are the first to investigate the robustness of backdoor countermeasures with initial concentrating on three influential model inspection based ones, which is (inadvertently) neglected by the community but is imperative for backdoor detection. For each countermeasure, two non-robust cases are disclosed.
    \item For the Neural Cleanse~\cite{wang2019neural}, it is formally proven to be non-applicable to binary-classification models and empirically shown to be ineffective in detecting backdoor in complex models especially when the network is deeper (e.g., ResNet 101).
    \item For the ABS~\cite{liu2019abs}, it fails in detecting a backdoor attack with a common poisoning data rate (e.g., 11\% essentially relatively high) where the attack success rate is nearly 100\%. It also fails in the other case where the weights of a backdoored model are constrained with small perturbations. It appears that the ABS assumption of backdoor effect being reflected by only one or a few compromised neurons can rarely be held.
    \item For the MNTD~\cite{xu2021detecting}, it is shown to be non-applicable to detecting backdoor for a given \textit{single} model, a common case for a given task even although it does exhibit good area under curve (AUC, a metric used in MNTD~\cite{xu2021detecting}). 
    It is further demonstrated to be sensitive to its underling hyperparameters as it uses meta-classifier, and fails once the hyperparameter slightly deviates even its AUC metric is used.
    \item For the three model inspections, the MNTD appears to be the least robust and most computing intensive. The Neural Cleanse is the most robust in that it can work in common cases under its threat model.
\end{itemize}

The rest of paper is organized as follows. Section~\ref{sec:related} provides related works on backdoor attacks and major differences between adaptive attack and non-robust cases of the countermeasures. Section~\ref{sec:deadspace} introduces the methodology of each focused countermeasure and identifies their non-robust cases. We conclude this work in Section~\ref{sec:conclusion} with a call on evaluating the robustness of other influential DL backdoor countermeasures in the future.
\section{Related Works}\label{sec:related}
\vspace{-0.1cm}
\subsection{Backdoor Attack}~\label{sec:backdoorAttack}
When a deep learning (DL) model is backdoored, the backdoored model $F_{\Theta}^{bd}$ behaves normally when it is fed with normal inputs without containing the attacker secretly chosen/known trigger, and the inference can be denoted as $ z=F_{\Theta}^{bd}(x)$ with $x$ as a clean input and $z$ as its ground-truth label. 
In this case, 
the inference accuracy of the backdoored model is indistinguishable from the accuracy of a clean model counterpart ($F_{\Theta}$).
Therefore, by merely validating the inference accuracy with a held-out validation dataset without knowing the secret trigger, it is infeasible to identify the insidious behavior of the backdoored model.
However, once a trigger $t$ is stamped on an input, the trigger input $x_t$ will be classified into the attacker set targeted label by the backdoored model, achieving the attacking purpose. 
Significantly, the attacking successful rate of the backdoor is usually very high, 100\% or nearly to it. In addition, the trigger can be \textit{arbitrary}, thus enabling natural objects (i.e. eye-glass~\cite{wenger2021backdoor}, and T-shirt~\cite{ma2022dangerous}) as triggers to be robust in physical world.

 
Backdoor attacks against DL were first highlighted in  2017~\cite{gu2017badnets,chen2017targeted}. Since then, it has received extreme attention from both academic and industry due to its potential disastrous consequences and realistic attack scenarios. The attack can be stealthily resulted from data collection, model training outsourcing, and collaborative learning (e.g., federated learning)~\cite{gao2020backdoor,shejwalkar2021back}. 
Considering the severe consequences of backdoor attacks, great efforts have been devoted to detecting or eliminating backdoors both from either offline inspection~\cite{wang2019neural,chen2019deepinspect,tran2018spectral,chen2018detecting,tang2021demon} or online inspection~\cite{gao2019strip,doan2020februus,liu2019abs}.

\vspace{2pt}\noindent

\subsection{Backdoor and Trigger Types}\label{sec:type}

\subsubsection{Backdoor Type} 
Backdoor attacks have different backdoor types and most existing defenses focus on defending against source-agnostic backdoor attacks. In such an attack, any input regardless of its source class containing the trigger will fire the backdoor inserted in an infected model. There are other backdoor types and the representative one is source-specific backdoor attack. The backdoor is activated not only when the trigger is embedded within the input but also when input is selected from attacker-chosen source classes. If the input is from a non-source class, the backdoor does not exhibit even though the input is with the trigger. The source-specific backdoor attack is challenging~\cite{tang2021demon} to defeat as most countermeasures including~\cite{wang2019neural,liu2019abs,gao2019strip} are ineffective against it.

\subsubsection{Trigger Type}  
Most defenses have relatively strong assumptions on the trigger types, such as size and pattern, which may not be held in practice. For example, the trigger size could be overlaid with the entire image input with certain transparency, which trivially makes many defenses ineffective~\cite{wang2019neural,doan2020februus}. Instead of targeting a label with a single trigger, multiple triggers can be used to target the same label, which is beyond the assumption of~\cite{liu2019abs} that cannot be detected. In addition, Neural Cleanse is less effective when the trigger pattern becomes complicated~\cite{guo2020towards} or network is deeper~\cite{ma2021quantization} or belongs to the feature-space based trigger attack~\cite{liu2019abs}.

Nonetheless, to explicitly demonstrate the non-robust cases of existing countermeasures, the backdoor type (i.e. source-agnostic backdoor) and trigger type (a small fixed patch trigger) used in this work are exactly within the threat model of each countermeasure. We do not consider any backdoor or trigger out the scope of the countermeasures in this work.

\subsection{Adaptive Attack}
In the area of AI security race such as adversarial example attack, model stealing attack, and this newly backdoor attack, it is unsurprising that the adaptive attack is extremely challenge to be defeated~\cite{gao2020backdoor,hu2021artificial}. The adaptive means that the attacker has knowledge of the defense strategy, thus developing more intelligent attacks to bypass it. In the adversarial example attack, it has been always shown that the new defenses are venerable to adaptive attacks~\cite{carlini2019evaluating,carlini2017adversarial}, similarly in backdoor defenses~\cite{shokri2020bypassing,nguyen2020input}. For instance, Shokri \textit{et al.}~\cite{shokri2020bypassing} devise an adaptive adversarial training algorithm to optimize the original loss function of the model, and also maximizes the indistinguishability of the latent representations of trigger input and benign input to bypass backdoor detections. For adaptive attacks, once it is developed, it's effectiveness is independent of e.g., dataset, task when being incorporated to bypass the defenses.

This work focuses on the non-robust cases of the backdoor countermeasures, the non-robust cases raise mainly because of the countermeasure's unknown sensitivity to e.g., i) dataset, task, model architecture of a potentially backdoored model or ii) hyperparameter of the countermeasure itself. It is not delicately designed adaptive attacks.

\section{non-robust cases Revelation}\label{sec:deadspace}
We firstly describe the used datasets that are MNIST~\cite{lecun1998gradient}, CIFAR10~\cite{krizhevsky2009learning} and GTSRB~\cite{stallkamp2012man}, since they have been used by the three focused model inspection based backdoor countermeasures. For each countermeasure, we firstly summary its inspection methodologies, which we follow when reproducing its source code. We then reveal and elaborate on two evasive cases serving as non-robust cases \textit{per countermeasure}. 

\subsection{Experimental Setup}
\subsubsection{Dataset}
\noindent{\bf MNIST} is a dataset consisting of 10 categories of handwritten digits provided by different people. It has 60,000 training images and 10,000 test images, both of which are $28\times 28 \times 1$ grayscale images. \textit{The MNIST has been used in Neural Cleanse and MNTD}.

\noindent{\bf CIFAR10} is a natural color image dataset for object recognition. It consists of 10 categories, and each category has 6,000 $32\times 32 \times 3$ RGB images---60,000 in total. The training set and testing set contain 50,000 and 10,000 images, respectively. \textit{The CIFAR10 has been used in ABS and MNTD}.

\noindent{\bf GTSRB} is a dataset for German traffic sign recognition (GTSRB), which contains 43 types of traffic signs. In the data preprocessing stage, all traffic signs are cut out from the image according to the bounding box coordinates, and aligned uniformly into a $32\times 32\times 3$ RGB image---the original size of the image varies. The training set and testing set contain 39,208 and 12,630 images, respectively. \textit{The GTSRB has been used in Neural Cleanse and ABS}.

\subsubsection{Trigger}\label{sec:trigger}
The trigger of the CIFAR10 is set to be the white square at the bottom right of the image. The trigger of GTSRB is set to the white triangle at the bottom left corner. For both datasets, the first category is selected as the target category that means any input stamped with the trigger will be classified into the first category by the backdoored model---so that we consider the source-agnostic backdoor type that is under threat model of each countermeasure. It is worth noting that the size of the two triggers accounts merely 3.5\% of the whole image, which are again within the effective detection range of the threat model of all three countermeasures. These two datasets along with the triggers are used to evaluate the Neural Cleanse and ABS.

As for the MNTD, it uses CIFAR10 and MNIST for evaluation. The trigger pattern and target category are randomly generated from the general attack setting distribution modeled by jumbo learning according to the source code of MNTD~\cite{xu2021detecting}.


\subsubsection{Metrics}
We employ two commonly used metrics to evaluate the performance of backdoored models: clean data accuracy (CDA) and attack success rate (ASR).

\begin{itemize}
\item {\bf CDA} refers to the percentage of clean test samples \textit{without triggers} that are correctly predicted to be their ground truth labels.

\item {\bf ASR} refers to the percentage of infected test samples \textit{containing the attacker chosen trigger} that are predicted to be the attacker-targeted label.
\end{itemize}

\subsubsection{Machine Configuration}
Our test machine is MECHREVO with NVIDIA GeForce RTX 3070 GPU (8\,GB video memory), Intel i7-11800H CPU (16 logical cores) and 16\,GB DRAM memory. All experiments including model training and defence evaluation are done on this machine.

\subsection{Neural Cleanse}

\subsubsection{Methodologies}
Neural Cleanse stems from the key intuition: the backdoor trigger can cross the decision boundary of the model, akin to create a ``shortcut" from the ground-truth label to the target label in the latent space 
of the trigger sample~\cite{wang2019neural}. Neural Cleanse estimates the minimum perturbation increment required to misclassify samples of \textit{any} other categories into a certain category, thus generating the smallest ``shortcut" as the potential reverse-engineered trigger for this category.
The specific implementation processes of Neural Cleanse are divided into three steps.

\begin{itemize}
\item Step 1: Reverse-engineer a smallest trigger/perturbation for a given label by using the optimization algorithm.

\item Step 2: Perform step one iteratively to generate a one-to-one corresponding reverse-engineered trigger per label.

\item Step 3: Estimate whether there is a reverse-engineered trigger of a category is \textit{significantly} smaller than other categories. The significant outlier reverse-engineered trigger means a real trigger, and the label that matches the trigger is the attacker targeted label\footnote{Note if such an outlier is not found, the Neural Cleanse will deem the model as benign and will not output infected label.}.
\end{itemize}

After producing all reverse triggers in Step 3, Neural Cleanse uses the Median Absolute Deviation (MAD) to estimate the anomaly index of the model. This method is more flexible in processing outliers than standard deviations, and can greatly reduce the impact of outliers on assessment robustness. More specifically, it calculates the $L_1$ norm of all reverse-engineered triggers, where the reverse-engineered trigger of the suspected infected label has a substantially small $L_1$ norm value, standing for the smallest perturbation. Then the median of absolute deviations between each norm and the median is calculated and recorded as MAD. The anomaly index is defined as the absolute deviation of the minimum norm divided by the MAD. Finally, a constant estimator (1.4826) is used to normalize the anomaly index. The anomaly index $AI$ is expressed as:
\begin{equation}\label{eq:anomaly index function}
\begin{aligned}
AI &= \frac{\lvert \textsf{Min}(L_{1}) - \textsf{Med}(L_{1}) \lvert}
{\textsf{Med}(\lvert L_{1} - \textsf{Med}(L_{1}) \lvert) \times C}
\end{aligned}
\end{equation}
where $L_1$ is the $L_1$ norm of all reverse-engineered triggers and $C$ is the constant estimator. The $\textsf{Med}(\cdot)$ and $\textsf{Min}(\cdot)$ represent the median function and the minimum function, respectively. Any model with an anomaly index greater than 2 is more than 95\% likely to be a backdoored model.

\subsubsection{Non-Robust Cases}

\paragraph{Case 1: Binary Classification}
Neural Cleanse is not well suit for a classification task with only few classes. In particular, for the binary classification, we prove that the anomaly index of the target model is always equal to 0.67, so that the Neural Cleanse is completely non-applicable in this case.

Binary classification has only positive and negative samples in the dataset. Neural Cleanse reverse-engineers a trigger for positive and negative sample, respectively. Assuming that the $L_1$ norms of reverse-engineered trigger 1 and reverse-engineered trigger 2 are $t_{1}$ and $t_{2}$, respectively, and $t_{1}$\textless $t_{2}$. According to \autoref{eq:anomaly index function}, the anomaly index can be expressed as:
\begin{equation}
\begin{aligned}
AI &= \frac{\lvert \textsf{Min}([t_{1}, t_{2}]) - \textsf{Med}([t_{1}, t_{2}]) \lvert}
{\textsf{Med}(\lvert [t_{1}, t_{2}] - \textsf{Med}([t_{1}, t_{2}]) \lvert) \times C} \\
&= \frac{- t_{1} + (t_{1} + t_{2}) / 2}
{[- t_{1} + (t_{1} + t_{2}) / 2 + t_{2} - (t_{1} + t_{2}) / 2] / 2  \times C} \\
&=\frac{1}{C}, 
\end{aligned}
\end{equation}
and vice versa for $t_{1}$\textgreater $t_{2}$. Among them, the $C$ is set to  1.482, so the anomaly index calculated by Neural Cleanse for the binary classification problem will be constantly equal to 0.67.

This $AI$ value is much smaller than the threshold of 2. Therefore, Neural Cleanse  will consider any 2-class model as a benign model regardless whether it is essentially backdoored or not.

\observ{The Neural Cleanse is non-applicable to binary classification tasks as the anomaly index is proven to be a constant of 0.67}

\begin{table}[h]
\centering 
\caption{Parameters of the four complex model architectures and their various metrics.}
\resizebox{0.45 \textwidth}{!}
{
\begin{tabular}{c | c | c | c | c}
\toprule 
\multirow{3} * {Model}& \multirow{3} * { layers} & \multirow{3} * {Params}
&\multicolumn{2}{c}{CDA(\%) of $M_{cl}$} \\
&&&\multicolumn{2}{c}{CDA(\%) $\&$ ASR(\%) of $M_{bd}$} \\
\cline{4-5}
&&&CIFAR10 & GTSRB \\ 

\hline
\multirow{2} * {VGG19}
& \multirow{2} * {19} & \multirow{2} * {39.0M}
& 92.70 & 98.84 \\
&&& 92.70 $\&$ 99.82 & 98.95 $\&$ 99.87 \\

\hline
\multirow{2} * {ResNet101}
& \multirow{2} * {101} & \multirow{2} * {42.5M}
& 95.08 & 99.02 \\
&&& 95.01 $\&$ 99.88 & 98.47 $\&$ 99.62 \\

\hline
\multirow{2} * {Inceptionv4}
& \multirow{2} * {150} & \multirow{2} * {41.2M}
& 93.15 & 99.07 \\
&&& 93.24 $\&$ 99.73 & 98.69 $\&$ 99.35 \\

\hline
\multirow{2} * {ResNeXt152}
& \multirow{2} * {152} & \multirow{2} * {33.2M}
& 95.48 & 99.26 \\
&&& 95.31 $\&$ 99.78 & 98.90  $\&$  99.4 \\
\bottomrule
\end{tabular}
}
\begin{tablenotes}
\small
\item{}
\end{tablenotes}
\label{tab:Complex model metrics}
\end{table}

\begin{figure}[h]
	\centering
	\includegraphics[trim=0 0 0 0,clip,width=0.45 \textwidth]{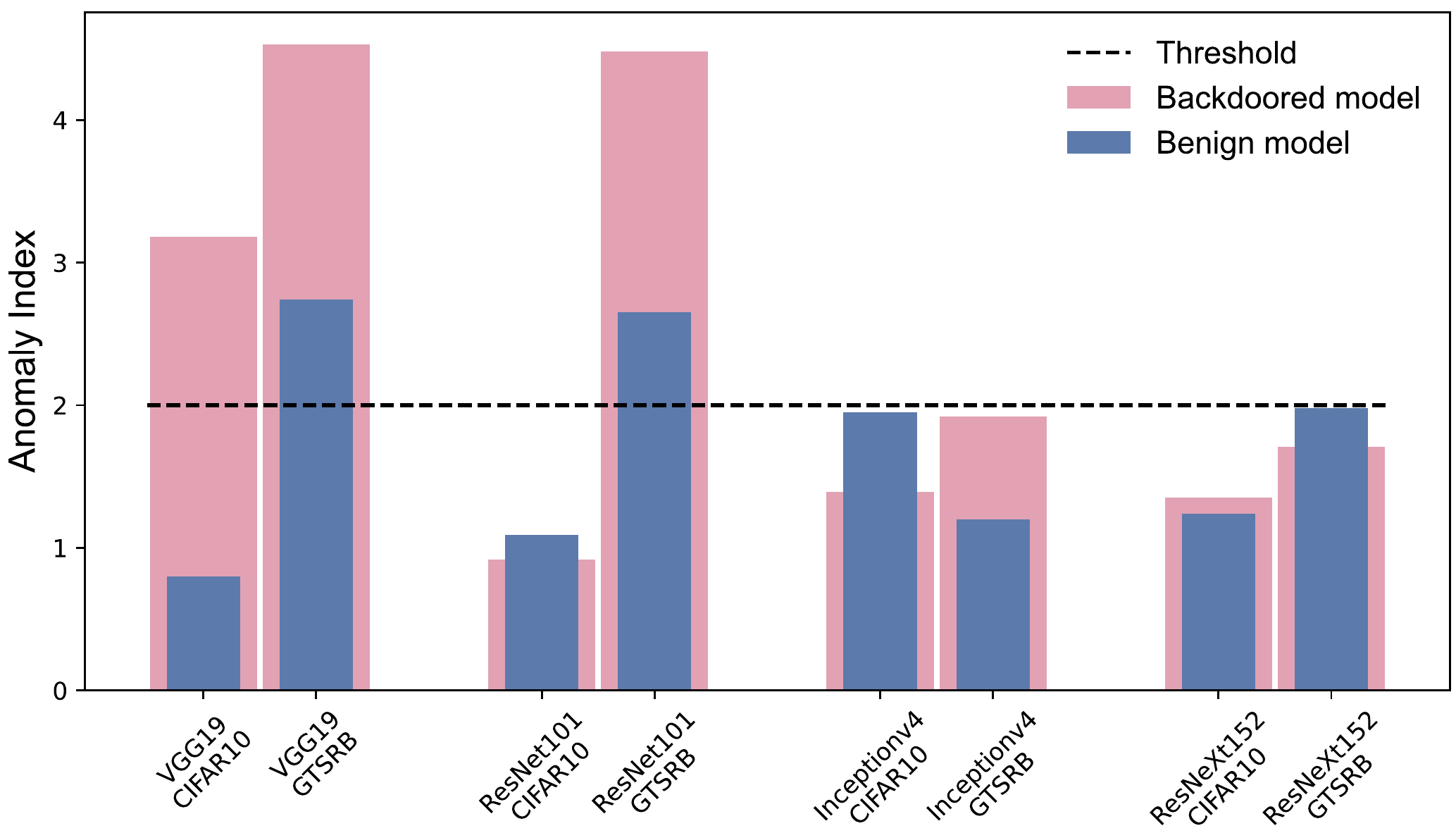}
	\caption{Neural Cleanse performance on four complicated model architectures.}
	\label{fig:AI}
\end{figure}

\begin{figure}[h]
	\centering
	\includegraphics[trim=0 0 0 0,clip,width=0.45 \textwidth]{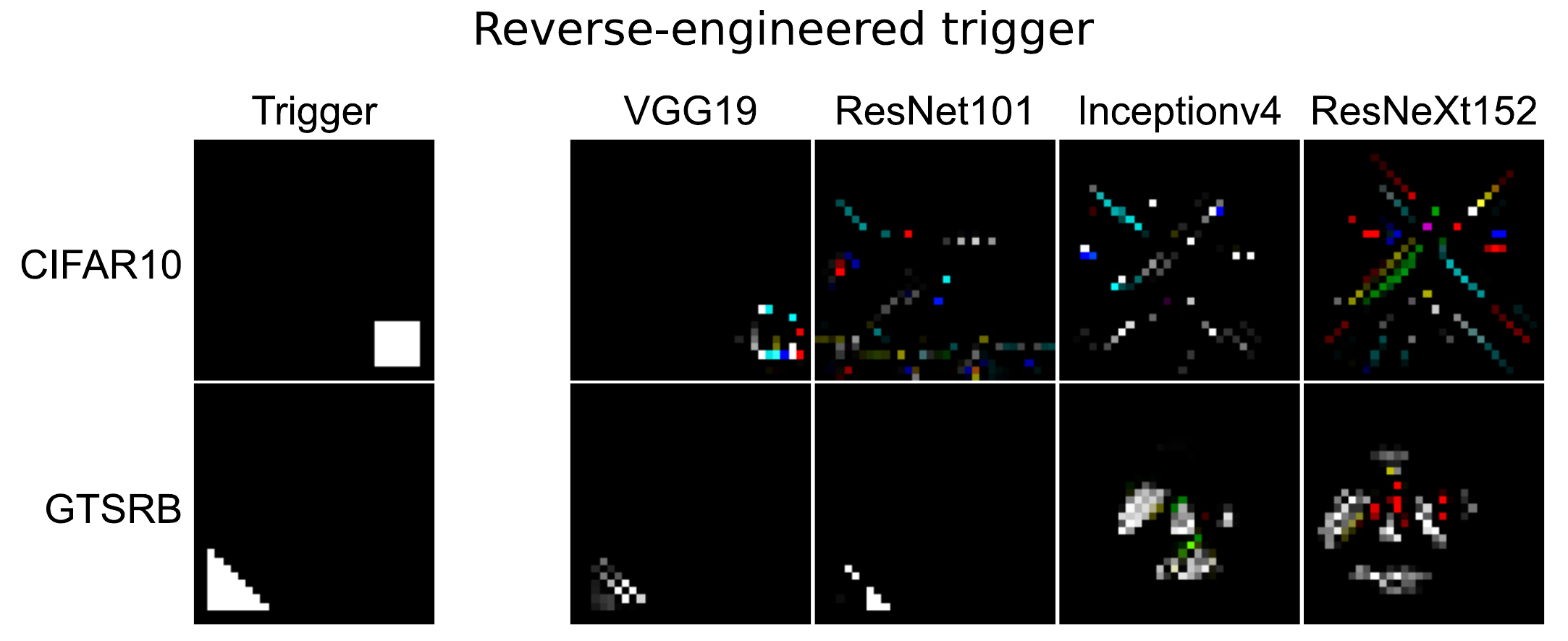}
	\caption{The trigger reverse-engineered by Neural Cleanse on the target label. Notably, the intentionally used small and simple trigger here should be easily and correctly reverse-engineered by the Neural Cleanse in principle.}
	\label{fig:NCTrigger}
\end{figure}

\paragraph{Case 2: Complicated Model}
Neural Cleanse is not good at detecting more complex model architectures, such as a deeper network that are with more parameters~\footnote{The source code of Neural Cleanse is at \url{https://github.com/bolunwang/backdoor}, and our reproduction is based on this source code.}.






For this case, we consider four models with much higher number of parameters that are suitable for recognition tasks: VGG19~\cite{simonyan2014very}, ResNet101~\cite{he2016deep}, Inceptionv4~\cite{szegedy2017inception}, and ResNeXt152~\cite{xie2017aggregated}. The trigger follows the settings in Section~\ref{sec:trigger}.
A clean model and a backdoored model are trained for each model architecture on each task of CIFAR10 and GTSRB. Note that VGG19 acts as a comparative role. Although it has more parameters, VGG19 model is far shallower than other models. The number of parameters, number of layers of each of four models are detailed in \autoref{tab:Complex model metrics}. This Table also shows that the clean model and its backdoored model counterpart have almost same CDA while the backdoored model exhibits nearly 100\% ASR.

Anomaly indices $AI$ for clean and backdoored model of each of four model architectures on both CIFAR10 and GTSRB are detailed in \autoref{fig:AI}. The Neural Cleanse fails in all cases except the VGG19+CIFAR10. More specifically, the Neural Cleanse always regards VGG19+GTSRB and ResNet101+GTSRB as backdoored models as their $AI$s are above the threshold of 2.0. While for the rest cases, their $AI$s are below 2.0, indicating that the Neural Cleanse regard them all as benign models. Overall, it is noted that the $AI$ gap between the clean model and the backdoored model becomes small as the model goes deeper, see \autoref{fig:AI}.

The potential failure reason could be that the feature vector extracted by the model of complex structure is more abstract, which may cause a certain degree of difficulty to the reverse engineering process of Neural Cleanse. As shown in \autoref{fig:NCTrigger}, as the model becomes more complex (in particularly, deeper), the reverse-engineered trigger tends to present the characteristics of the targeted category samples rather than that of the trigger. The reverse-engineered trigger deviates more from the original trigger as the model is more complex.
As $AI$ greater than the threshold of 2 means that the backdoored model is identified by the Neural Cleanse and vice versa, the backdoored model with deeper network cannot be detected---always regarded as benign (shown in \autoref{tab:Complex model metrics}).

\observ{The Neural Cleanse almost always fails for deeper complicated networks. In addition, the behaviors (i.e. $AI$) detected by Neural Cleanse between the clean model and the backdoored becomes less distinguishable when the model goes deeper}






\subsection{ABS}
\subsubsection{Methodologies}
Artificial Brain Stimulation (ABS) is inspired by electrical brain stimulation (EBS), which uses a controllable way to change the activation value of an individual artificial neuron but fix the rest to study whether it is damaged~\cite{liu2019abs}. The ABS is based on two key observations: one is that there are damaged neurons in the backdoored model, and the backdoor effect is likely to be dominated by one or a group of neurons
in the model. Another observation is that the subspace of the target label in the decision space can cross the decision boundary of all labels, and the activation of damaged neurons can cause the model to classify the trigger sample into the target label.

\vspace{0.2cm}
\noindent\textit{Neuron Stimulation Function (NSF) of each Neuron.} In the process of identifying whether the model has a backdoor or not, ABS explores the influence of internal neurons' activation on the output activation of each label. 
Specifically, ABS observes the forward inference process of the model on benign samples and fixes the activation values of other neurons belonging to the same layer as one neuron $\alpha$ to be tested. Immediately afterwards, the neuron $\alpha$ is stimulated to change its activation value to calculate its effect on the activation of subsequent layers.
Then it stimulates the neuron $\alpha$ to change its activation in order to calculate its effect on the activation of the subsequent layers. Finally, the relationship between the activation the neuron $\alpha$ and the activation of the output layer is determined, which is called neuron stimulation function (NSF).

\vspace{0.2cm}
\noindent\textit{Damaged Neurons Identification.} After the above process, the NSF of each neuron in certain chosen layers
is calculated (testing all neurons in all layers could be costly). If the activation of a certain neuron within a certain range  \textit{significantly increases the output activation of a specific label}, the neuron is marked as a damaged neuron candidate. ABS can mark the 10 most likely damaged candidate neurons through the NSF of the neuron.

\vspace{0.20cm}
\noindent\textit{Backdoored Model Determination.} After filtering out the damaged candidate neurons in the model, an optimized method is leveraged to reverse engineer a trigger per candidate neuron.
If the reversed trigger is effective for other benign samples, then the neuron is affirmed to be a damaged neuron and the trigger is the final reversed trigger. Following the settings in the ABS released code, it can be determined that the model has a backdoor when the reversed trigger based attack success rate (REASR) is greater than 88\%\footnote{This threshold of 88\% is following the setting in the source code.} 

\subsubsection{Non-Robust Cases}

\paragraph{High ASR with Low Poison Rate} 
In~\cite{liu2019abs}, the ASR of the backdoored model under evaluation and the poison rate is not clearly given when reporting the REASR score\footnote{We note that in Table~2 of~\cite{liu2019abs}, it does consider three poison rate of 1\%, 9\% and 50\%. However, only a single REASR score is reported that is always 100\% or nearly 100\%, it is unclear whether this score refers to the best case or other case---only the best case matches our reevaluations.}.
In our reproduction experiments, we noticed that the ABS always fails to detect the backdoor even it exhibits nearly 100\% ASR---same as reported in~\cite{liu2019abs} but the backdoor is implanted with a not very high poison rate (i.e. 11\%).

Following the experimental settings~\cite{liu2019abs}\footnote{The source code is at \url{https://github.com/naiyeleo/ABS}, which we reproduce.}, the model is backdoored through data poisoning~\cite{gu2017badnets}. Specifically, given the training dataset, a subset is randomly selected and the trigger is stamped on each sample, then the sample label is modified to be the targeted label. The fraction of the subset to the entire set is the poison rate. We use a set of different poisoning rates: 50\%, 11\%, 1.5\%  and 0\% (clean model). Four backdoored models are trained based on each setting, where the model architectures are VGG16 and ResNet18: each trains on CIFAR10 and GTSRB.
The backdoor attack type is the class-agnostic attack that ABS aims to detect. The trigger follows the settings in Section~\ref{sec:trigger}.

\begin{figure}[h]
	\centering
	\includegraphics[trim=0 0 0 0,clip,width=0.45 \textwidth]{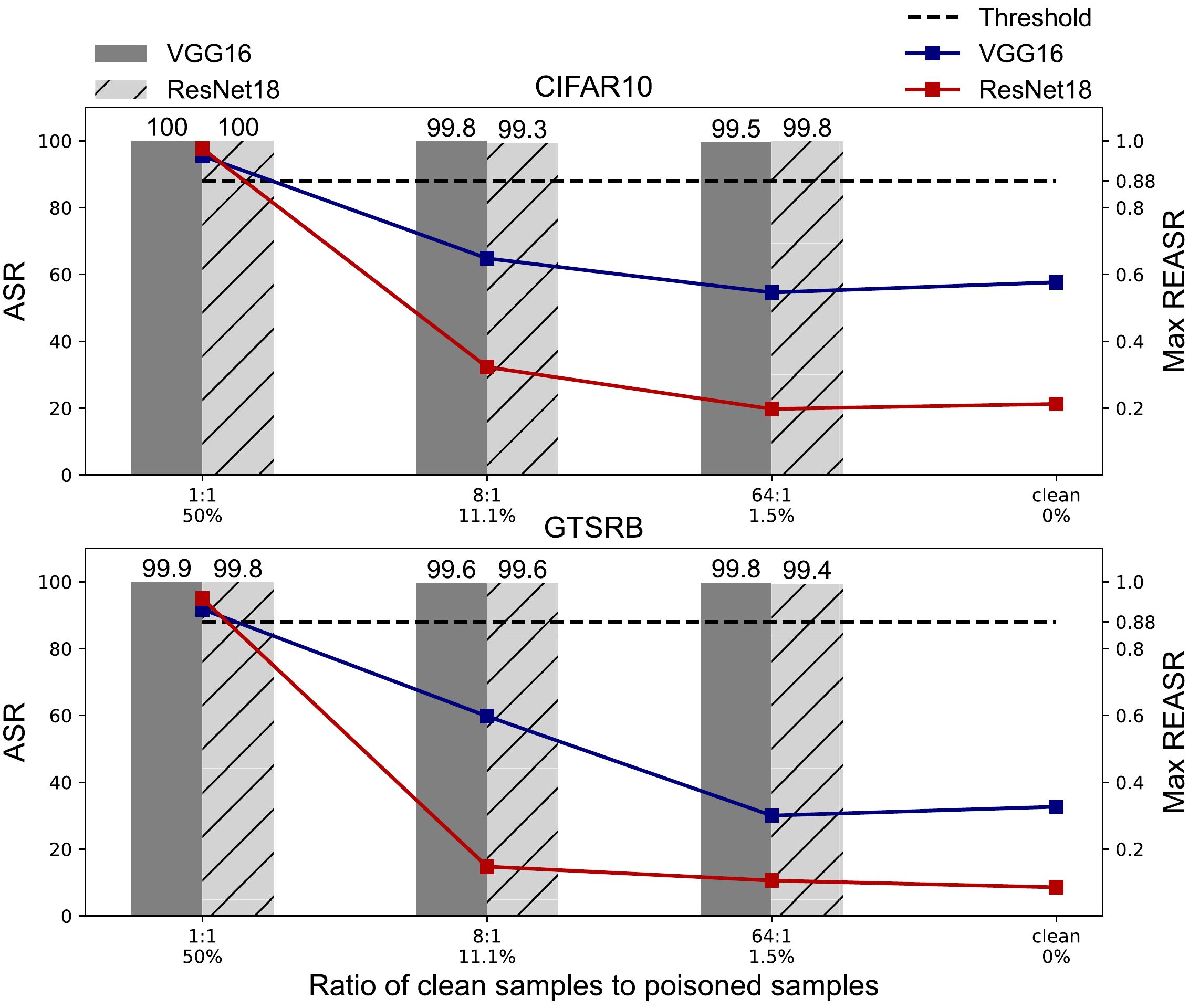}
	\caption{ASR of the backdoored model and the REASR detection score by the ABS. The 0\% poison rate corresponds to a clean model.}
	\label{fig:reasr}
\end{figure}

The ASR and REASR of backdoored models are detailed in \autoref{fig:reasr}. Although the backdoored model still has a very high attack success rate (over 99\%), ABS does not consider it a backdoored model even the poison rate is up to 11\%. This is because the REASR score of the ABS is below than the threshold of 0.88. One possible reason is that the training set with a higher rate of poisoning samples (i.e. 50\%) will affect the behavior of neurons more drastically during the process of injecting the backdoor, causing the backdoor effect to be expressed by only one or a few of neurons. 
On the contrary, a training set with a not very high poisoning rate will trivially cause less damage to a given specific neuron, and its malicious behavior is scattered throughout the neural network, requiring a large number of neurons to be cooperated for the sake of effective backdoor.

It is notable that the data poisoning based attack usually relies on a small poisoning rate, e.g., several percent or even below 1\%~\cite{gao2021design}. As in \autoref{fig:reasr}, we can observe that the REASRs of 11.1\%, 1.5\% and 0\% poison rates are quite similar. Note the 0\% poison rate corresponds to the clean model, this therefore means that the REASRs of 11.1\%, 1.5\% poison rates are almost indinstinguishable from the clean model. Therefore, the ABS appears to be quite cumbersome in the reality. It is also noted that the REASR score of the ResNet18 is much smaller than that of the VGG16 for both CIFAR10 and GTSRB datasets. This implies that the ABS is sensitive to model architecture. Moreover, according to \autoref{fig:reasr}, the REASR is also dependent on the dataset: GTSRB has a lower value than CIFAR10.

\observ{The ABS is unable to detect backdoored model even if the ASR is nearly 100\% when the poisoning rate for implanting the backdoor is not very high (in fact, the failure case of 11\% poisoning rate is already high). In addition, the ABS appears to be sensitive to model architecture and dataset}

\paragraph{Adversarial Weight Perturbation}: 
We have shown above that the ABS can trivially fail. The main reason lies on the fact that it has a almost unrealistic \st{strong} assumption about the backdoored model: the backdoor effect is expressed by one or a group of neurons. The ABS building upon on such assumption can be easily bypassed by, even if the backdoored attack type is still (class-agnostic attack) and the ASR is exceptionally within its threat model.

\begin{table}[h]
\centering 
\caption{The Max REASR of adversarial weight perturbation enabled backdoor attacks.}
\resizebox{0.45 \textwidth}{!}
{
\begin{tabular}{c | c | c | c} %
\toprule 
Task & Model  &
CDA(\%) $\&$ ASR(\%)   &
\begin{tabular}{@{}c@{}} Max REASR of \\ $M_{cl}\,|\, M_{bd}$ \end{tabular} \\ 

\midrule
\multirow{2}*{CIFAR10} 
& ResNet18 & 93.01 $\&$ 99.84 & 0.2124 $|$ 0.2176 \\
& VGG16 & 92.32 $\&$ 99.82 & 0.5767 $|$ 0.5603 \\

\midrule
\multirow{2}*{GTSRB} 
& ResNet18 & 97.85 $\&$ 99.78 & 0.0856 $|$ 0.0958 \\
& VGG16 & 97.55 $\&$ 99.91 & 0.3268 $|$  0.3116\\
\bottomrule
\end{tabular}
}
\begin{tablenotes}
\small
\item The $M_{\rm cl}$ and $M_{\rm bd}$ are short for clean model and backdoored model, respectively.
\end{tablenotes}
\label{tab:reasr}
\end{table}

We next consider the other case. We use an adversarial weight backdoor attack method inspired by~\cite{garg2020can}, which utilizes projection gradient descent (PGD) to constrain the weight of the backdoored model within the $l_{\infty}$ error range of the weight of the benign model. The projection operator is formulated as  $P_{l_{\infty}}(\Theta_{M}, \varepsilon)$, where $\Theta_{M}$ represents the parameter of the benign model, and $\varepsilon$ represents the constraint range and is set to be as small as 0.01 (this value can be varied to be larger and does not affect our conclusion). In other words, we first train a clean model, then the backdoored model is fine-tuned based on the clean model using poisoned samples, where the fine-tuning is guided by the PGD to only inject small perturbations to the neurons of the clean model. Attributing to the mandatory constraints of PGD, the backdoored effect are reflected by a large number of neurons, thus easily bypassing the strong assumption of ABS.

As for experiments, the model architecture uses VGG16 and ResNet18: each trains on CIFAR10 and GTSRB, respectively. The selection of the trigger is the same as above, the poisoning sample rate is set to 0.5, that is, half of the training samples are infected. Note such a poisoning rate is indeed abnormally high for the source-agnostic backdoor attack. The 0.5 poisoning rate is to make sure the ABS can detect the backdoored model. In this way, the strength of adversarial weight perturbation based backdoor attacks on ABS can be explicitly demonstrated. The final model performance metrics in terms of the CDA, ASR and REASR score of the ABS are detailed in \autoref{tab:reasr}. Firstly, the REASR of the backdoored model is below the detection threshold, the backdoored model cannot be detected even the ASR is still nearly 100\%. Secondly, the REASR of the backdoored model is nearly same to the clean model. Thirdly, the REASR score again is smaller for deeper network even for the clean model, indicating the ABS sensitivity to model architecture. 

\observ{The strong assumption of the ABS that is the backdoor effect dominated by a single or a few neurons makes it quite fragile. It is neither effective against the not-high poisoning rate based backdoor or the weight perturbation based backdoor attack even that the ASR is (nearly) 100\%. Generally, such an assumption is rarely held in practice}

\subsection{MNTD}
\subsubsection{Methodologies}
Meta Neural Trojan Detection (MNTD) aims to train a meta-classifier that can recognize backdoored models~\cite{xu2021detecting}. The meta-classifier takes metadata extracted from the model to be detected as input and performs binary classification to determine whether it has a backdoor or not. Given a target model that is going to be inspected and a small-scale clean dataset, the MNTD follows three steps.

\begin{figure}[t]
	\centering
	\includegraphics[trim=0 0 0 0,clip,width=0.4 \textwidth]{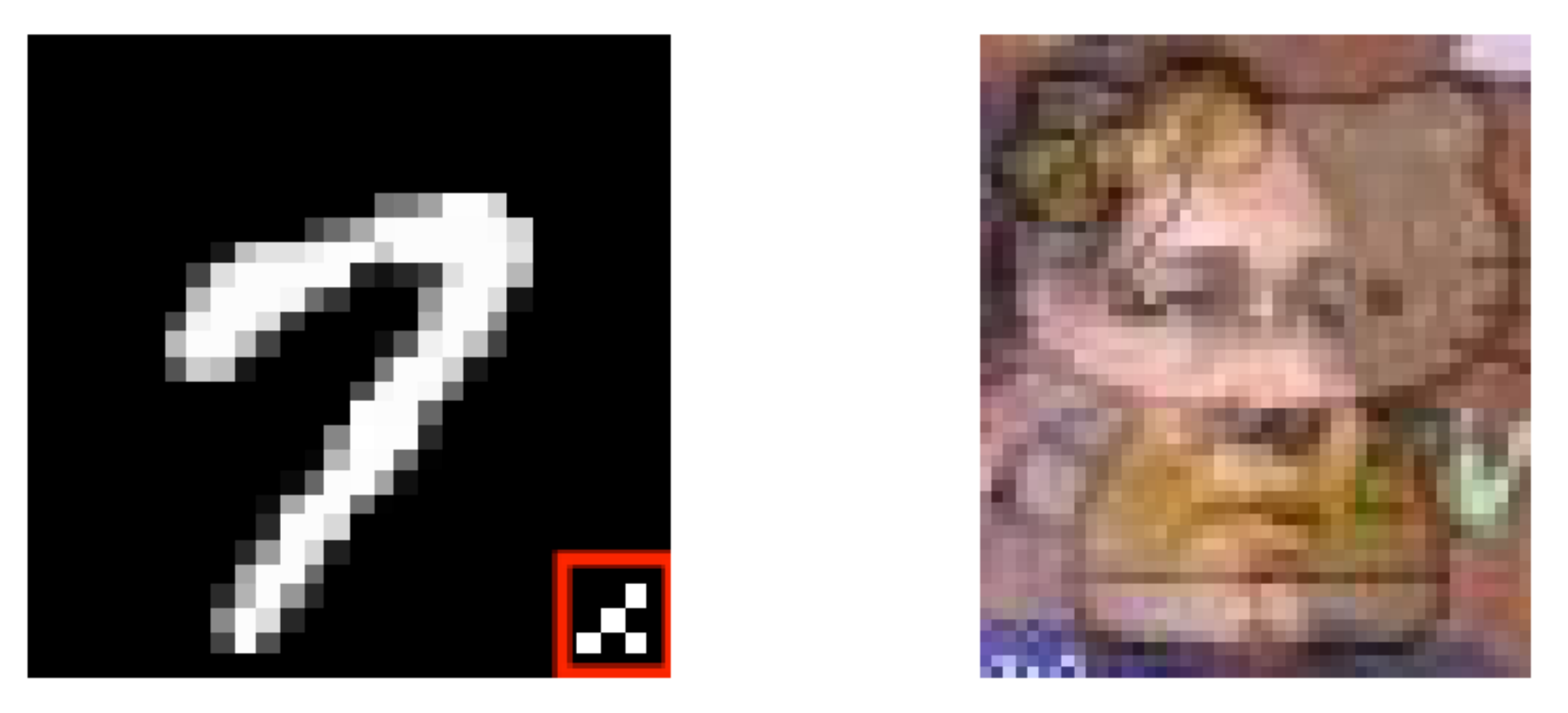}
	\caption{The image on the left shows a trigger sample of a modification backdoor attack, the red box in the bottom right corner highlights the trigger; the image on the right shows a trigger sample of a blending backdoor attack, the trigger is a translucent Hello Kitty doodle.}
	\label{fig:M_B}
\end{figure}

First, a large number of shadow models (i.e. several thousands~\cite{xu2021detecting}) is generated as the meta dataset for training and testing the meta-classifier. More specifically, benign models are trained with different parameter initializations on a clean dataset as positive samples. In addition, jumbo learning is leveraged to generate multiple backdoor attack settings, and then train several backdoored models per different setting as negative samples. It is worth noting that the two main types of attacks in MNTD's backdoor attack settings include modification attacks~\cite{gu2017badnets} and blending attacks~\cite{chen2017targeted}. A modification attack directly modifies some part of the sample as a trigger and sets the label of the poisoned sample as the target label. In contrast, a blending attack blends the trigger pattern into the original sample (e.g. blending a specific pattern into the image). The former is usually a small size trigger while the later is normally a large size trigger but transparent, as exemplified in \autoref{fig:M_B}.

Secondly, the meta-classifier is trained on the above prepared benign and backdoored shadow models. Generally, a small number of samples/queries are fed into each shadow model, which gives outputs (in particular, logits according to its source code implementation\footnote{In the description of~\cite{xu2021detecting}, it refers the output as representation vector, but does not explicitly mention what it is.}). These outputs (logits) serve as metadata are inputs of the meta-classifier (the outputs from different queries are concatenated, 
while benign or backdoored labels are the ground-truth labels of the meta-classifier. To be efficient, a number of queries are optimized during the training of the meta-classifier and will be used during model inspection. 

Finally, a well-trained meta-classifier is used to detect the target model. The optimized queries are used as the input of the target model to obtain the logits. The meta-classifier will receive the logits as input and output a score. The score is used to measure the malicious degree of the target model, and the expected score of the clean model is smaller than that of the backdoored model. The meta-classifier takes the median of the scores of all shadow models in the training set as the threshold. If the score of the target model is greater than the threshold, it indicates that it is injected into the backdoor.
\subsubsection{Non-Robust Cases}

\paragraph{Caveats of AUC as Detection Metric} 
The MNTD uses Area Under Curve (AUC) as the primary metric for evaluating the meta-classifier performance. AUC usually needs to be computed on a set of benign and backdoored models trained on the same dataset over which the targeted model trains, and the computation process is independent of a threshold. In practice, the detection \textit{has to be performed per target model}, which often relies on a threshold to determine whether the given target model has a backdoor or not. However, it is unclear how to choose such a threshold in~\cite{xu2021detecting}, thus determining whether a given model is backdoored or not\footnote{In Section II.B of~\cite{xu2021detecting} when describing the workflow of meta neural analysis, it does mention the inference is done on a given target model, but there is no evaluations on a single model in its experiments.}.

According to the settings in the source code\footnote{\url{https://github.com/AI-secure/Meta-Nerual-Trojan-Detection}}, the median of all shadow model scores in the \textit{testing set} is chosen as the threshold. However, this threshold determination appears to be unreasonable.
In general, the model trainer should be assumed to have no threshold knowledge of the test set unless it is already tested.
Therefore, we instead choose the median of all shadow model scores in the \textit{training set} as the threshold, rather than the test set.

Even in this context, we find that the correct threshold chosen is still challenging.
Because there is a significant difference between the thresholds in the training set and the thresholds in the testing set. In other words, the usage of the training set determined threshold is non-applicable to separating positive and negative samples (benign and backdoored models) in the testing set, which indicates that it is challenging to adopt the MNTD in practice.

In \autoref{tab:threshold}, determined thresholds from five trained meta-classifiers\footnote{The source code has trained five meta-classifiers that we follow.}
are detailed. In addition, for each meta-classifier, we use the threshold of the training set to evaluate the shadow models in the testing set (testing set consists of 256 benign and backdoored models, respectively, same as the source code setting), 
which accuracy results are also detailed. The accuracy is the ratio that the number of models correctly judged by the meta-classifier---whether the targeted model is benign or backdoored---to the number of all models under evaluation. The targeted model is backdoored if the score is higher than the threshold, otherwise, benign. We can see that the accuracy is always 50\% akin to guessing. In addition, we can see that there exists a large difference among the training set determined thresholds even the meta-classifiers are trained on the same task (i.e. MNIST). In addition, the thresholds between the training and testing sets given the same meta-classifier are also diverse.

\observ{The MNTD is challenging to inspect model individually as the AUC metric cannot do so, while the median score of meta-classifier as threshold exhibits an almost guessing performance. So it is unclear how to apply the MNTD to inspect an individual model per task that are more likely demanded in practice}

\begin{table}
\centering 
\caption{Threshold and accuracy per meta-classifier for MNIST and CIFAR10}
\resizebox{0.45 \textwidth}{!}
{
\begin{tabular}{c | c | c | c | c} %
\toprule 
\multirow{2} * {Dataset} &
\multirow{2} * {\begin{tabular}{@{}c@{}} Number of \\ meta-classifier \end{tabular}} &
\multirow{2} * {\begin{tabular}{@{}c@{}} Threshold of \\ training set \end{tabular}}  &
\multicolumn{2}{c}{Threshold $\&$ Accuracy(\%)} \\
\cline{4-5}&&&
Test set-M & Test set-B \\

\midrule
\multirow{5} * {MNIST}
& 1 & -4.69 & 15.70 $\&$ 50.00 & 17.39 $\&$ 50.00 \\
& 2 & -0.80 & 6.93 $\&$ 50.20 & 7.70 $\&$ 50.19 \\
& 3 & -0.36 & 8.33 $\&$ 50.00 & 9.62 $\&$ 50.00 \\
& 4 & -3.46 & 5.60 $\&$ 50.00 & 7.03 $\&$ 50.00 \\
& 5 & -4.39 & 1.60 $\&$ 50.00 & 2.42 $\&$ 50.00 \\

\midrule
\multirow{5} * {CIAFR10}
& 1 & -2.64 & 2.98 $\&$ 50.00 & 3.15 $\&$ 50.00 \\
& 2 & -0.24 & 3.14 $\&$ 50.00 & 3.21 $\&$ 50.19 \\
& 3 & 0.03 & 8.09 $\&$ 52.34 & 7.89 $\&$ 52.54 \\
& 4 & -2.90 & 8.57 $\&$ 50.00 & 8.14 $\&$ 49.80 \\
& 5 & -3.26 & 7.64 $\&$ 50.00 & 7.24 $\&$ 50.00 \\

\bottomrule
\end{tabular}
}
\begin{tablenotes}
\small
\item -M means the backdoored model is attacked by a modification attack, while -B means a blending attack.
\end{tablenotes}
\label{tab:threshold}
\end{table}

\paragraph{Sensitivity to Hyperparameters} As shown in \autoref{tab:threshold}, there is a significant variance between the thresholds of the meta-classifier on the training set and on the testing set. This may be caused by the difference in hyperparameters used to train the shadow models during the training and testing phases. In this case, we further explore the effect of hyperparameters on the MNTD detection performance of the testing set and confirm that the meta-classifier is highly sensitive to hyperparameters.


We investigate hypterparameters including i) epochs, ii) batch sizes and iii) the proportion of the training samples (i.e. from CIFAR10) used to train the shadow models in the MNTD. The original hyperparameters used in the source code of~\cite{xu2021detecting} are summarized in \autoref{tab:original hyperparameters}.
For each hyperparameter, we perform several different sets of control experiments to train shadow models used for testing. In other words, for the hyperparameters to train shadow models used to train the meta-classifier, they are fixed to be same as the source code (as in \autoref{tab:original hyperparameters}). In the process of generating the test shadow models, we change only one of these three hyperparameters and kept the other two fixed to investigate the explicit effect of a single hyperparameter on the MNTD detection performance.

\begin{table}
\centering 
\caption{Original hyperparameters the MNTD provided in the source code.}
\resizebox{0.35 \textwidth}{!}
{
\begin{tabular}{c | c | c | c} %
\toprule 
Dataset & Epochs & Batch size & Proportion \\

\midrule
Training set
& 100 & 100 & 0.02 \\

Test set
& 4 & 100 & 0.5 \\

\bottomrule
\end{tabular}
}
\label{tab:original  hyperparameters}
\end{table}

\begin{table}
\centering 
\caption{The metrics of the shadow models trained using the original MNTD source code settings (all these performance values are averaged result).}
\resizebox{0.35 \textwidth}{!}
{
\begin{tabular}{c | c | c | c | c} %
\toprule 
\multirow{2} * {Dataset} & \multicolumn{2}{c |}{Training Dataset} & \multicolumn{2}{c}{Test Dataset} \\
\cline{2-5}
& CDA(\%) & ASR(\%) & CDA(\%) & ASR(\%) \\

\midrule
MNIST & 95.14 & - & 98.50 & - \\

\hline
MNIST-M & \multirow{2} * {94.19} & \multirow{2} * {86.06} & 98.35 & 99.72 \\

\cline{1-1} \cline{4-5}
MNIST-B & & & 98.25 & 98.56 \\

\midrule
CIFAR10 & 39.33 & - & 61.36 & - \\

\hline
CIFAR10-M & \multirow{2} * {38.40} & \multirow{2} * {82.12} & 61.30 & 99.73 \\

\cline{1-1} \cline{4-5}
CIFAR10-B & & & 59.18 & 88.70 \\

\bottomrule
\end{tabular}
}
\label{tab:shadow  metrics}
\end{table}

\begin{figure}[t]
	\centering
	\includegraphics[trim=0 0 0 0,clip,width=0.4 \textwidth]{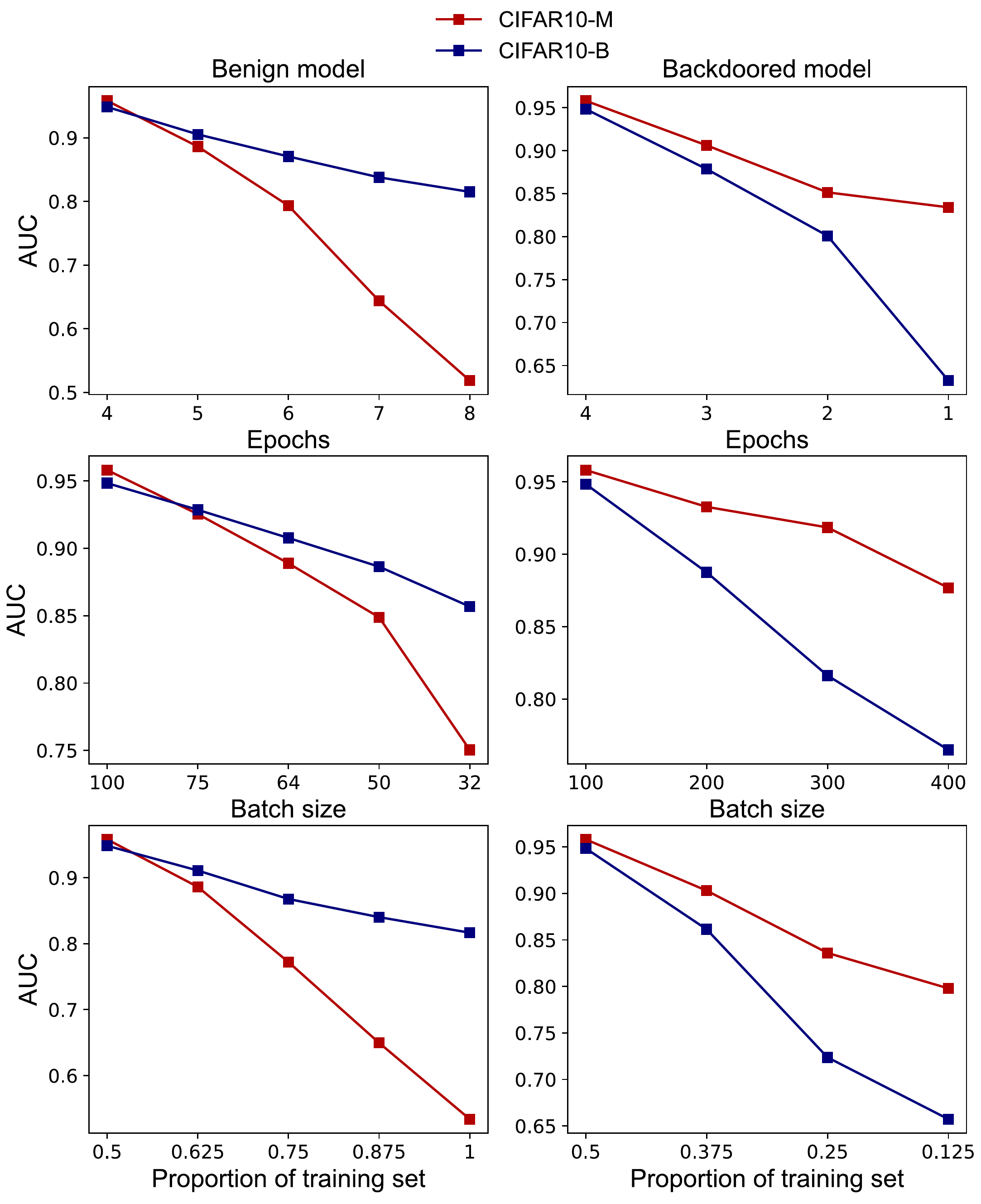}
	\caption{The influence of hyperparameters on AUC. (top) number of epochs, (middle) batch size, (bottom) proportion of training set.}
	\label{fig:AUC}
\end{figure}

For each controlled hyperparameter, we train a set of test shadow models consisting of 256 benign ones and 256 backdoored models on the CIFAR10 dataset. The model architecture is kept same with the one used in the source code (i.e. 4 convolutional layers + 3 fully connected layers). For the training of meta-classifier, it is same as the source code as well---we do not change.
After meta-classifier is trained, it is employed to evaluate the test shadow models corresponding to each controlled hyperparameter. For the detection performed, we use the AUC to be same as~\cite{xu2021detecting} to \textit{solely reflect} the influence of hyperparameter on the MNTD. 
\autoref{tab:shadow  metrics} summarizes the shadow benign/backdoored models averaged performance by reproducing the MNTD according to the same setting in \autoref{tab:original  hyperparameters}. To reduce computational overhead, the MNTD trains shadow models without fully converge---even such partial training is already extremely costly, e.g., 14 hours to complete the 4096 shadow models training (though 12 seconds per shadow model) in the CIFAR10 case even on the (4 conv + 3 full) shallow network. 
As a comparison, training the well converged model takes only about e.g., one hour for ResNet50. This somehow means that the MNTD is inapplicable to the backdoor induced in the common model training outsource scenario. As it is more reasonable for the user to train the model by herself/himself given that the MNTD does require heavy computation and ML expertise---costing more than training from scratch. The AUCs are detailed in \autoref{fig:AUC} for the specific change of epochs, batch size, and proportion of training set, respectively. 

Before diving into detailed analyses, we provide necessary knowledge about how the MNTD meta-classifier scores the benign and backdoored models to ease the following understandings. According to the source code, the backdoored models are generally scored higher in the meta-classifier than benign models under the \textit{same hyperparameter settings}. Consequentially, the evaluated AUC exhibit a higher value (i.e. close to 100\%) by virtue of the larger distribution difference between the two scores of backdoored and benign models. In other words, if the difference in the distribution of scores between the two decreases, then the AUC will degrade (i.e. close to 50\%). 

\begin{table*}
\centering 
\caption{Metrics and MNTD evaluation results for VGG11-based test sets.}
\resizebox{0.9 \textwidth}{!}
{
\begin{tabular}{c | c | c | c | c | c | c} %
\toprule 
Model & Epochs & Batch size & Proportion & CDA(\%) $\&$ ASR(\%) & Median of scores & AUC(\%)\\

\midrule
Benign & 50 & 100 & 1 & 80.02 $\&$ \textover[c]{-}{99.99} & 46.83 & \multirow{2} * {43.20} \\
Backdoored & 50 & 100 & 1 & 79.84 $\&$ 99.99 & 45.18 \\

\midrule
Benign & 100 & 100 & 1 & 80.65 $\&$ \textover[c]{-}{99.99} & 82.86 & \multirow{2} * {13.30} \\
Backdoored & 50 & 100 & 1 & 79.84 $\&$ 99.99 & 45.18 \\

\bottomrule
\end{tabular}
}
\label{tab:vgg_shadow}
\end{table*}

\vspace{0.2cm}
\noindent\textit{Epochs:} The number of epoch when training the test benign models is varied from 4 to 8 (the source code uses 4)---the batch size and proportion of training set are fixed to be 100 and 0.5, respectively. As shown in Fig.~\ref{fig:AUC} (top row), the AUC decreases as the epoch number goes up. For the CIFAR10-M, the AUC almost degrades to 50\% for epoch number of 8. 
In contrast, when the number of epochs of test backdoored models decreases from 4 to 1 (the source code uses 4), the AUC also degrades e.g., to 65\% in the CIFAR10-B. It appears the closer the benign model converges (more epochs), the higher score for it by the meta-classifier. 
As the score of backdoored models are fixed, the score difference between the benign and backdoored models is closer, resulting a lower AUC. Similarly, when less closer the backdoored model converges, the lower the score given by the meta-classifier. This also results in a smaller score gap between the benign and backdoored models, thus a smaller AUC.

\vspace{0.20cm}
\noindent\textit{Batch Size:} In this setting, the batch size when training the test benign and backdoored models varies while keep the rest two being constant (4 and 0.5 for epochs and proportation of training set, respectively). As can be seen from \autoref{fig:AUC} (middle row), the AUC of benign models notably decreases as the batch size is reduced from 100 to 32. The AUC decreases when the batch size of the backdoored model increases from 100 to 400. Note a smaller batch size means there are more interactions in a given epoch, or the model will be converged faster. 
As the meta-classifier scores higher for the more converged models, we can observe the smaller score difference between benign models and backdoored models when the benign models converge better with higher score by the meta-classifier and the backdoored models' higher score fixed. Similarly, when the benign model's lower score is fixed and the backdoored models are less converged resulting in lower score, the score difference between the beign and backdoored models are also decreases, leading to a degraded AUC.

\vspace{0.20cm}
\noindent\textit{Proportion of Training Set:} In this setting, the samples used for training the test benign and backdoored models are changed while the rest two being fixed (4 and 100 for epochs and batch size, respectively). The AUC results are shown in \autoref{fig:AUC} (bottom row). When test benign models are trained with more samples, it converges better, thus higher scored by the meta-classifier. This consequently leads to smaller score difference between the benign models and backdoored models that are fixed with higher scores. So that a lower AUC is exhibited. Similarly, less training samples for backdoored models result in lower score by meta-classifier, smaller score difference, consequently degraded AUC.

Based on above analyses, we can see that either a larger epoch or smaller batch size, or more training data (a higher proportion of the training set) will result in the benign model being more fully trained and closer to convergence, at which point the meta-classification will output a higher score for it. Conversely, the less adequately trained the backdoored model is, the lower the model score will be. Both cases lead to a reduction in the difference between the score distribution of the benign model and the backdoored model, resulting in a lower AUC. Therefore, the tendency for backdoored models to score higher than benign models is likely to be caused by the fact that the \textit{backdoor sub-task} in the backdoored model always converges (i.e. high ASR) rather than the presence of a backdoor in the model itself, aslo validated in \autoref{tab:shadow  metrics}.





\begin{figure}[t]
	\centering
	\includegraphics[trim=0 0 0 0,clip,width=0.4 \textwidth]{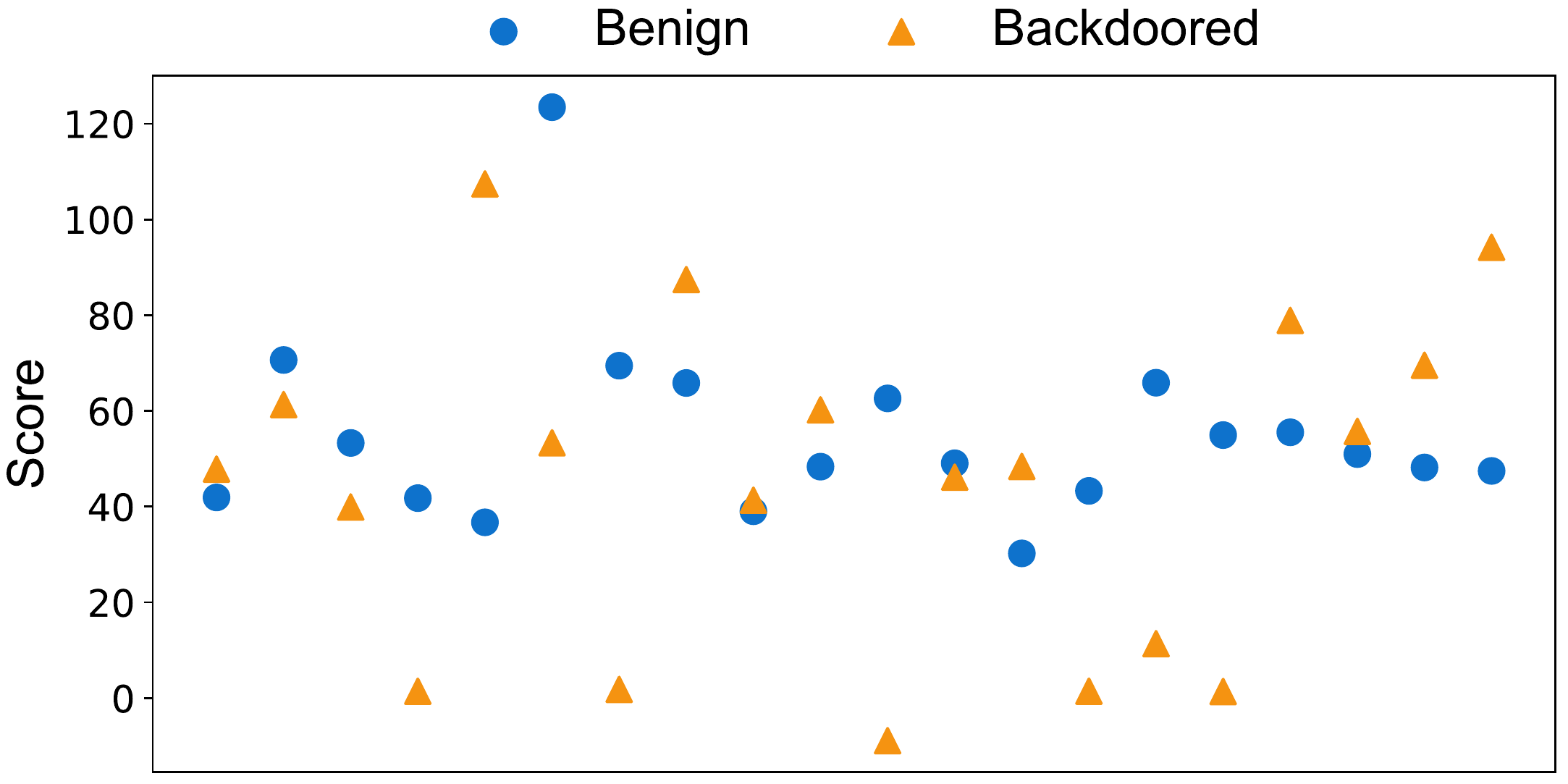}
	\caption{Score distribution of the VGG11-based test set, extracted from one of the meta-classifiers.}
	\label{fig:vgg_score}
\end{figure}

Above experiments follows the same setting of the MNTD code with same model architecture when training training shadow models and testing shadow models. The training shadow models are usually not fully converged---doing so would incur further substantial training overhead. While for the target models under inspection, it is more reasonable that the target models are converged well and using a different model architecture.
In this context, we simulate a more real scenario that both the testing benign and backdoor models have converged. We followed the training procedure for the shadow model and trained 40 VGG11 models to convergence on the CIFAR10 dataset as the test set, with half the positive and half the negative samples.
The training parameters and performance on the meta-classifier for the VGG11-based test set are shown in \autoref{tab:vgg_shadow}.
As both the benign and backdoor models were trained to convergence, the meta-classifier gave high scores for both. When the training parameters are the same regardless of the presence of a backdoor in the model, the distribution of scores for the converged model is similar to that shown in \autoref{fig:vgg_score}, with an average AUC of close to 50\% for the MNTD evaluation (2nd and 3rd rows in \autoref{fig:vgg_score}). And when the training epoch for the benign model is increased to 100 on top of this, the average AUC decreases further (4th and 5th rows in \autoref{fig:vgg_score}). Notably, when we examine the score median for the backdoored model and benign models, the score of the benign models is surprisingly overtaken the backdoored models.

In summary, the meta-classifier is extremely sensitive to hyperparameters that are used to train its shadow models and the targeted models, where a slight change (i.e. epochs variation) is sufficient to render the meta-classifier fail to identify the backdoored model. For instance, the AUC decreases from 96\% to 50\% when the epoch of the test benign models is increased from 4 to 8 in the CIFAR10-M test set---50\% is simply guessing. In other words, the meta-classifier seems to prefer to give higher scores for the well-trained models (i.e. those model converged well) rather than only for the backdoored models.

\observ{The MNTD is found to be (highly) sensitive to hyperparameters used to train the model that is under inspection, which renders its usage difficulty even when the AUC as a detection metric is used. Because the AUC is sensitive when the hyperparameter (i.e. epochs, batch size) varies. Note that the MNTD is (at least inexplicitly claimed) generic to different backdoor types (i.e. source-agnostic backdoor) and triggers. It however easily fails for the simplest backdoor attack and easiest-to-detect triggers}

\section{Conclusion}\label{sec:conclusion}
This work opens a new research area of critically examining the non-robust cases of DL backdoor countermeasures, which is inadvertently \textit{neglected} to date. The root cause why a countermeasure is non-robust is from its inherent characteristics, e.g., datasets, tasks, model architectures and even its hyperparameters, which are unexplored and not elucidated. No clarification of these non-robust cases will give normal user false confidence after applying those defenses. For example, the Neural Cleanse falsely regards backdoored model with deeper network always as benign, while the ABS falsely deems the backdoored model to be benign if the infected model is trained with a normal poison rate (i.e. a small poison rate is a common practice). Significantly, it is found that the MNTD, as an AI-against-AI defense, is (highly) sensitive to hyperparameters, which tends to easily provide false security indications that should be carefully adopted.
We have identified two non-robust cases per each most influential existing model inspection based backdoor countermeasures to initially demonstrate the urgency of understanding their often neglected non-robust cases. We hope more efforts will be devoted to this research area in the future or the emerging devised countermeasures should critically examine and clarify their non-robust cases more thoroughly.

\bibliographystyle{IEEEtran}
\bibliography{References}

\end{document}